\documentclass[aps,prl,twocolumn,notitlepage,showpacs,superscriptaddress,10pt]{revtex4-2}%
\usepackage{graphicx}
\usepackage{amsmath}
\usepackage{amssymb}
\usepackage{color}
\usepackage{amsfonts}%
\usepackage[caption=false]{subfig}

\usepackage{mathbbol}
\usepackage{color}
\usepackage[dvipsnames]{xcolor}
\usepackage[colorlinks,bookmarks=false,citecolor=darkblue,linkcolor=red,urlcolor=blue]{hyperref}
\usepackage[capitalize]{cleveref}

\definecolor{darkred}{rgb}{0.7,0.0,0.0}

\definecolor{darkblue}{rgb}{0,0.02,0.45}

\definecolor{darkgreen}{rgb}{0.02,0.45,0.0}

\definecolor{violet}{rgb}{0.8,0.2,0.6}

\providecommand{\U}[1]{\protect\rule{.1in}{.1in}}

\newcommand{\bicup}{Bi$_2$Sr$_2$CuO$_6$}

\begin{document}
\title{Altermagnetism from a Cu-Fe Lieb Lattice in FeSe/Cuprate Heterostructures}

\author{Ying Li}
\affiliation{MOE Key Laboratory for Nonequilibrium Synthesis and Modulation of Condensed Matter, School of Physics, Xi'an Jiaotong University, Xi'an 710049, China}
\affiliation{Physics Department, Technical University of Munich, TUM School of Natural Sciences, 85748 Garching, Germany}

\author{Augustin Davignon}
\affiliation{D\'epartement de Physique et Institut Quantique, Universit\'e de Sherbrooke, Sherbrooke, J1K 2R1, Qu\'ebec, Canada.}

\author{Peng Rao}
\affiliation{Physics Department, Technical University of Munich, TUM School of Natural Sciences, 85748 Garching, Germany}
\affiliation{Munich Center for Quantum Science and Technology (MCQST), Schellingstr. 4, 80799 München, Germany}

\author{Runhan Li}
\affiliation{D\'epartement de Physique et Institut Quantique, Universit\'e de Sherbrooke, Sherbrooke, J1K 2R1, Qu\'ebec, Canada.}

\author{Maia G.~Vergniory}
\affiliation{D\'epartement de Physique et Institut Quantique, Universit\'e de Sherbrooke, Sherbrooke, J1K 2R1, Qu\'ebec, Canada.}
\affiliation{Donostia International Physics Center, Paseo Manuel de Lardizabal 4, 20018 Donostia-San Sebastian, Spain.}
\affiliation{Regroupement Qu\'eb\'ecois sur les Mat\'eriaux de Pointe (RQMP), Quebec H3T 3J7, Canada}

\author{Roser Valent{\'\i}}
\affiliation{Institut f\"ur Theoretische Physik, Goethe-Universit\"at Frankfurt,
Max-von-Laue-Strasse 1, 60438 Frankfurt am Main, Germany}

\author{Johannes Knolle}
\affiliation{Physics Department, Technical University of Munich, TUM School of Natural Sciences, 85748 Garching, Germany}
\affiliation{Munich Center for Quantum Science and Technology (MCQST), Schellingstr. 4, 80799 München, Germany}
\affiliation{Blackett Laboratory, Imperial College London, London SW7 2AZ, United Kingdom}

\date{\today}

\begin{abstract}
Realizing altermagnetism in high-$T_c$ cuprate-based systems would provide a direct route for studying spin-split electronic bands in the absence of net magnetization and investigate their interplay with unconventional superconductivity. Here, we propose that FeSe/cuprate heterostructures offer such a platform, where a 45$^\circ$ twist of Cu and Fe layers creates an effective CuFe\(_2\) Lieb lattice in which Fe magnetic order and Cu-Fe hybridization through the ligands induces altermagnetic \(d\)-wave spin splitting. A minimal tight-binding model shows that this mechanism is generic. Furthermore, a substrate-induced inequivalence of the two Se sites in FeSe provides a second route in which altermagnetism originates in the Fe layer and is transferred to the cuprate layer by proximity.
Density functional theory calculations for FeSe/\bicup~ heterostructures confirm the viability of both mechanisms and reveal ways to enhance the spin splitting. These results establish superconducting cuprate/transition metal chalcogenide heterostructures as a promising setting for engineering altermagnetism and studying its coupling to unconventional superconductivity.
\end{abstract}
\maketitle
\par
{\it Introduction.--}  Altermagnetism (AM) has been defined as a class of collinear magnetic order that combines the spin-split electronic bands of ferromagnets with the vanishing net moment of antiferromagnets~\cite{libor2020crystal, mazin2021prediction, libor2022emerging,libor2022beyond, libor2022giant, jungwirth2026symmetry, bai2024altermagnet, song2025altermagnet, mazin2024altermagnetism}, and provides a unified description of previously reported anomalous behavior in magnetic systems~\cite{ahn2019antiferromagnetism,hayami2019momentum, yuan2020giant, ma2021multi}. Various realizations of AM on lattice models~\cite{leeb2024spontaneous,das2024realizing,roig2024minimal,ferrari2024altermag,zeng2024bilayer,kaushal2026spontaneous} have been complemented by real-material counterparts that introduce additional richness beyond idealized geometries through strain, correlation, and other microscopic mechanisms~\cite{amin2024nanoscale, reimers2024direct, yang2025three, zhou2025manipulation, cui2026data, chakraborty2024strain, khodas2026tuning, peru2026altermagnetism, chen2026identifying,ahn2019antiferromagnetism, gao2025ai}. Among all candidates, the Lieb lattice stands out as the cleanest and most representative platform for hosting AM~\cite{li2025exploring, durrnagel2025altermagnetic,kaushal2025altermagnetism, ma2021multi, zhang2025crystal, jiang2025ametallic, garcia2026microscopic, wei2025la2o3mn2,Zhang2025giant, guo2026external, taskiran2026novel, liu2026uncompensated}. High-$T_c$ cuprates naturally realize an effective Lieb lattice once the ligand oxygen orbitals are explicitly included. 
The coexistence of superconductivity and AM enables exotic phenomena such as dissipation-free pure spin supercurrents~\cite{monkman2026persistent} and non-reciprocal Josephson transport~\cite{ouassou2023}, motivating the question of how AM can be engineered in superconducting Lieb-lattice systems.  In high-$T_c$ cuprates, AM has been proposed via inducing magnetic moments on oxygen sites~\cite{li2025exploring, fischer2011mean}, but this typically requires parameter regimes that appear unrealistic for material realization. This calls for the search of alternative, more robust routes to AM.
\begin{figure}[!htb]
\center
\includegraphics[width=\linewidth]{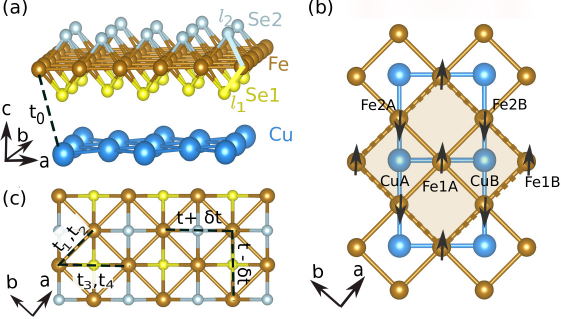}
\caption{(a) Side view of FeSe/Cu in the FeSe/cuprate heterostructure, where the yellow (light blue) atoms represent Se1 (Se2) located below (above) the Fe layers. The presence of the substrate induces a shorter Fe-Se1 distance compared to the Fe-Se2 distance. (b) Top view of the effective Lieb lattice CuFe$_2$ when omitting the ligands. (c) Top view of the FeSe lattice. $t_1$ and $t_2$ denote the nearest-neighbor hoppings, while $t_3$ and $t_4$ denote the next-nearest-neighbor hoppings. The distinct bond distances of Fe-Se1 and Fe-Se2 give rise to different next-nearest-neighbor Fe-Fe hoppings in a checkerboard pattern, denoted as $t + \delta t$ and $t - \delta t$. Orbital correspondences of these hoppings are provided in the Supplemental Material.}
\label{fig:model}
\end{figure}

In this letter, we propose a two-layer heterostructure to realize AM consisting of FeSe monolayers on CuO$_2$ stacks. Our proposal is based on the observation that the ratio between the Cu–Cu distance in the CuO$_2$ layer of high-$T_c$ cuprate superconductors ($d_{\rm Cu\text{–}Cu}\approx 3.8$~\AA~\cite{park1995Structures}) and the Fe–Fe distance in FeSe ($d_{\rm Fe\text{–}Fe}\approx 2.66$~\AA, corresponding to lattice constants 3.76 \AA~\cite{tan2013interface}) are close to $\sqrt{2}$.  Consequently, rotating the FeSe layer by 45$^\circ$ relative to the CuO$_2$ layer makes the combined CuFe$_2$ network form an effective close-to-ideal Lieb lattice, as shown in Figs.~\ref{fig:model} (a),  and (b). In fact, the two-dimensional Bi$_2$Sr$_2$CuO$_{6+\delta}$ (Bi-2201)~\cite{wang2023prominent} and Bi$_2$Sr$_2$CaCu$_2$O$_{8+\delta}$ (Bi-2212)~\cite{yu2019high} have recently attracted considerable interest because their $45^\circ$-twisted CuO$_2$ layers provide a promising platform for unconventional electronic properties~\cite{can2021high, margalit2022chiral, tummuru2022josephson, lee2021twisted, zhao2021emergent, zhu2021presence, li1999bicrystal, takano2002dlike, latyshev2004caxis, wang2023prominent}.
FeSe is also an ideal building block. It has the simplest structure among iron-based superconductors, exhibits strong magnetic tunability, and is highly sensitive to interface effects. For instance, monolayer FeSe on SrTiO$_3$ exhibits a dramatically enhanced critical temperature $T_C$ up to 65 K~\cite{huang2017monolayer, xu2021spect, ge2015super, he2013phase}.

There are two distinct mechanisms for realizing AM in the FeSe/cuprate-based 
heterostructure. The first route relies on the Lieb lattice, requiring a checkerboard AFM ordering on Fe atoms combined with a non-zero hybridization between Fe and Cu. The second route originates solely from the underlying heterostructure symmetries and proximity effects.
For instance,  in FeSe/SrTiO$_3$ heterostructures, the $\mathcal{P}\mathcal{T}$ symmetry in FeSe is broken by the SrTiO$_3$ substrate, enabling AM in FeSe~\cite{mazin2023induced}. In this specific case, AM arises due to the existence of two distinct next-nearest-neighbor hopping parameters between Fe atoms, which in turn stem from the non-equivalent Se1 (below) and Se2 (above) the Fe layer sites created by the presence of the substrate, as shown in Fig.~\ref{fig:model} (a). The AM order in FeSe can then be transferred to Cu by proximity effects.

Using effective tight-binding models and {\it ab initio} density functional theory (DFT) calculations, we demonstrate that both mechanisms cooperate under realistic conditions. To assess their individual roles, we investigate each mechanism separately. Our results establish FeSe/cuprate heterostructures as a tunable platform for realizing AM in the parent compounds  of high-$T_c$ superconductors and provide opportunities to explore its interplay with superconductivity.

{\it Effective Model.--} As shown in Fig.~\ref{fig:model} (a), the crystal structure of FeSe consists of Fe planes where Fe$^{2+}$ ions are in a $3d^6$ configuration, sandwiched between two square lattices of Se atoms. To model the electronic structure, we first treat the system in its simplest form using a two-orbital model on a two-dimensional square lattice of Fe atoms, retaining the two degenerate orbitals ($d_{xz}$ and $d_{yz}$) per Fe site following Ref.~\cite{raghu2008minimal}. 
On the other hand, the CuO$_2$ plane is characterized by Cu$^{2+}$ ions in a $3d^9$ configuration arranged on a square lattice, with a localized electron in the Cu $3d_{x^2-y^2}$ orbital. The interlayer hopping between Cu and Fe atoms is denoted as $t_0$.

To describe this system, we constructed an effective ten-band tight-binding model and incorporated the Hubbard $U$ term, which is treated at the mean-field level. The model is defined on a unit cell containing two clusters (labeled $L = A$ and $B$), with each cluster contributing a basis of one Cu and two Fe atoms [see Fig.~\ref{fig:model} (b)]: $(\text{Cu}_L, d^1_{L,xz}, d^1_{L,yz}, d^2_{L,xz}, d^2_{L,yz})$, where $d^i_{L,a}$ denotes the $d_a$ orbital on the $i$-th Fe atom in sublattice $L$. The explicit form of the Hamiltonian is given as:
\begin{align}
    h (\mathbf{k},\sigma) =& \begin{pmatrix} h_{AA}(\mathbf{k},\sigma) & h_{AB}(\mathbf{k},\sigma) \\ h_{AB}(\mathbf{k},\sigma)^\dagger & h_{BB}(\mathbf{k},\sigma)\end{pmatrix},
    \label{eq:10band1}
\end{align}
where
\begin{align}
      & h_{AA}(\mathbf{k}, \sigma) =\nonumber \\&
    \begin{pmatrix}
    \epsilon_{\text{Cu}} - \frac{U_{\text{Cu}}}{2}\sigma& t_0\gamma_{\text{CF}}^1 & -t_0\gamma_{\text{CF}}^1 & -t_0\gamma_{\text{CF}}^2 & -t_0\gamma_{\text{CF}}^2\\
 t_0\gamma_{\text{CF}}^{1*} & \epsilon_{\text{Fe}}^{xz}- \frac{U_{\text{Fe}}}{2}\sigma & 0 & t_2\gamma_{\text{F}}^1 & 0\\       -t_0\gamma_{\text{CF}}^{1*} & 0 & \epsilon_{\text{Fe}}^{yz}- \frac{U_{\text{Fe}}}{2}\sigma & 0&t_1 \gamma_{\text{F}}^1\\
       -t_0\gamma_{\text{CF}}^{2*}&t_2\gamma_{\text{F}}^1 & 0 &  \epsilon_{\text{Fe}}^{xz} + \frac{U_{\text{Fe}}}{2}\sigma & 0 \\
       -t_0\gamma_{\text{CF}}^{2*}& 0 & t_1 \gamma_{\text{F}}^1 & 0 &  \epsilon_{\text{Fe}}^{yz} + \frac{U_{\text{Fe}}}{2}\sigma
       \end{pmatrix}.
\end{align}
The expression of $h_{BB}(\vec{k}, \sigma)$ is the same as $h_{AA}(\vec{k}, \sigma)$, except that the upper left term is replaced by $\epsilon_{Cu} + (U_{\text{Cu}}/2)\sigma$. And $h_{AB}(\vec{k}, \sigma)$ is written as:
\begin{align}
    &h_{AB}(\mathbf{k}, \sigma) = \nonumber\\&
     \begin{pmatrix}    
   t_{\text{Cu}}\gamma_{\text{Cu}} & -t_0\gamma_{\text{CF}}^3 &t_0\gamma_{\text{CF}}^3 &t_0\gamma_{\text{CF}}^4 & t_0\gamma_{\text{CF}}^4\\   
   -t_0\gamma_{\text{CF}}^1 &h_{1}& h_{2}  &  t_1\gamma_{\text{F}}^2 & 0 \\ 
   t_0\gamma_{\text{CF}}^1 &h_{2}&h_{1}& 0 &  t_2\gamma_{\text{F}}^2 \\
   t_0\gamma_{\text{CF}}^2 & t_1\gamma_{\text{F}}^2 & 0 &h_{3}&h_{4}\\
   t_0\gamma_{\text{CF}}^2 & 0 & t_2\gamma_{\text{F}}^2 &h_{4}& h_{3}\\
    \end{pmatrix}.
    \label{eq:10band}
    \end{align}
    
The above terms in our model are organized as follows. Nearest-neighbor Fe-Fe hoppings are parameterized by $t_1$ and $t_2$ [see Fig.~\ref{fig:model} (c)]. The next-nearest-neighbor Fe-Fe hoppings, parameterized by $t_3$ and $t_4$ and mediated by Se atoms, enter the Hamiltonian through four terms denoted as $h_1$, $h_2$, $h_3$, and $h_4$. Their inequivalence along the two square diagonals arises because the Cu layer pushes the lower Se1 atoms toward the Fe plane, inducing a strain effect. This substrate-induced strain is captured by the coefficients $\delta t_3$ and $\delta t_4$, as shown in Fig.~\ref{fig:model} (c). We set $\delta t_3 = \delta t_4 = \delta t$ throughout this work. Detailed expressions of $h_1$-$h_4$,  $\gamma_F^{1,2}$,  and $\gamma_{CF}^{1-4}$ are provided in the Supplemental Material.

For the FeSe two-orbital model, we adopt the hopping parameters of Raghu \textit{et al.}~\cite{raghu2008minimal} ($t_1 = 1$ eV, $t_2 = -1.3$ eV, $t_3 = t_4 = 0.85$ eV). Since the purpose of these calculations is to elucidate the roles of $t_0$ and $\Delta t$ in the emergence of AM, we use this parametrization rather than aiming for a quantitative description of the FeSe band structure. For the Cu subsystem, we take $t_{\mathrm{Cu}} = -0.5$ eV, consistent with Ref.~\cite{sheshadri2023connecting}, with the chemical potential determined by the occupation number. The Hubbard interactions are set to $U_{\mathrm{Fe}} = 3$ eV and $U_{\mathrm{Cu}} = 4$ eV, following Refs.~\cite{Lanata2012,Watson2017,sheshadri2023connecting}. A comparison with a five-orbital DFT-based parametrization is provided in the Supplemental Material.

\begin{figure*}
\includegraphics[angle=0,width=\linewidth]{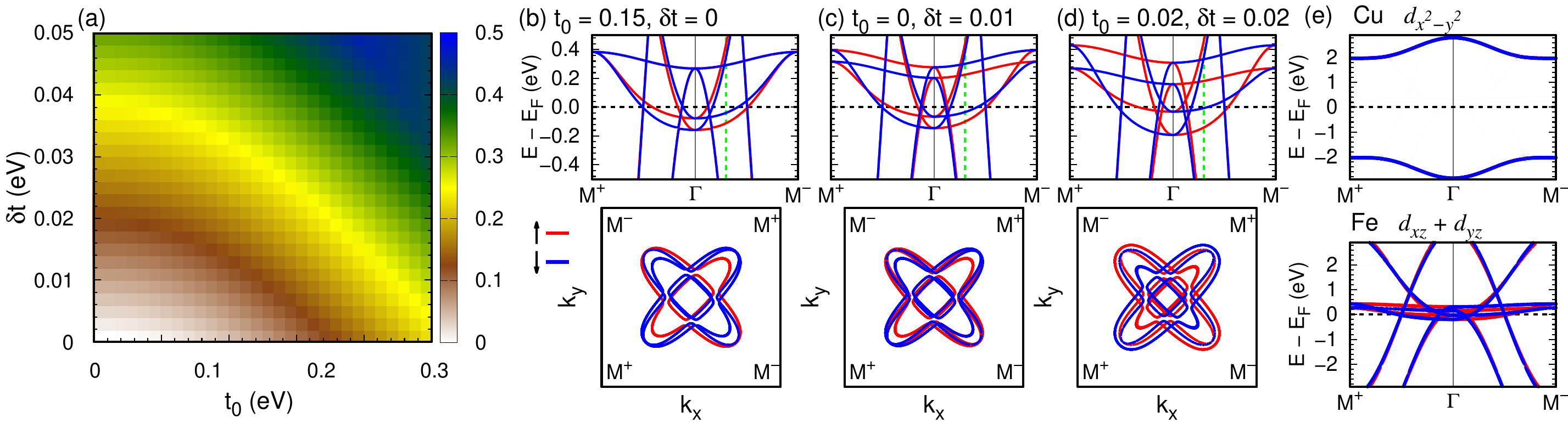}
\caption{Mean-field results of Eq.~(\ref{eq:10band1}), showing band splittings (eV) induced by $t_0$ (Cu-Fe interlayer hopping) and $\delta t$ (Fe-Fe hopping asymmetry) at the $K$ point (dotted green lines, top panels of b--d). Fermi surfaces are shown in bottom panels. Parameters: (b) the first mechanism only, with $t_0 = 0.15$ eV, $\delta t = 0$; (c) the second mechanism only, with $\delta t = 0.01$ eV, $t_0 = 0$; (d) parameters estimated for the realistic heterostructure, with $t_0 = 0.02$ eV and $\delta t = 0.02$ eV, where both mechanisms are present.
(e) Orbital-resolved spectral weights of the split bands, showing the Cu $d_{x^2-y^2}$ orbital (top) and the Fe $d_{xz}+d_{yz}$ orbitals (bottom).}
\label{fig:modelband}
\end{figure*}

\begin{figure}
\includegraphics[angle=0,width=\linewidth]{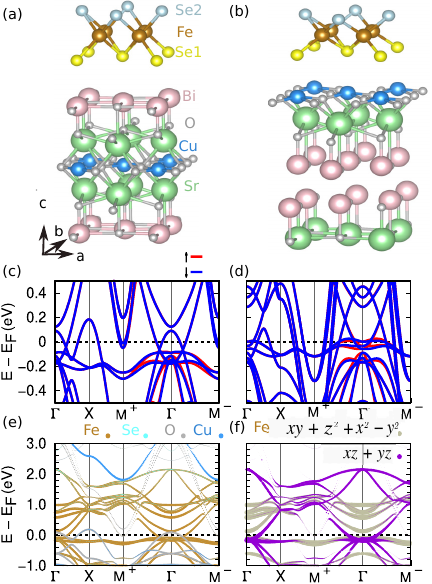}
\caption{Crystal structures of FeSe/\bicup~(FeSe/Bi-2201) heterostructures, where Bi-2201 is constructed from fabricating (a) a BiO$_2$ (structure A) and (b) CuO$_2$ planes (structure B). (c),(d) Corresponding band structures for structures A and B, respectively. (e),(f) Atom- and $d$-orbital-projected spectral weights of Fe for structure A, respectively.
}
\label{fig:DFT}
\end{figure}

The numerical results of the mean-field calculations are presented in Fig.~\ref{fig:modelband}. As introduced in the introduction, the two AM mechanisms are controlled by two key parameters: $t_0$, which denotes the hopping between Cu and Fe, and $\delta t$, which characterizes the substrate-induced inequivalence of next-nearest-neighbor Fe-Fe hoppings in FeSe. The band splittings as a function of $t_0$ and $\delta t$ are shown in Fig.~\ref{fig:modelband} (a). 
The corresponding band structures and the Fermi surfaces are displayed in Figs.~\ref{fig:modelband} (b) - (d). Fig.~\ref{fig:modelband} (b) corresponds to the first mechanism alone ($t_0 = 0.15$ eV, $\delta t = 0$), while Fig.~\ref{fig:modelband} (c) corresponds to the second mechanism alone ($\delta t = 0.01$ eV, $t_0 = 0$). Finally, Fig.~\ref{fig:modelband} (d) shows the band structure obtained using the DFT-estimated parameters ($t_0 = 0.02$ eV, $\delta t = 0.02$ eV)~\footnote{From the DFT results for structure B in Fig.~\ref{fig:DFT} (b), we determine the ratios $(t_0/t_2)^{\text{DFT}}$ and $(\delta t/t_2)^{\text{DFT}}$. Multiplying these ratios by the model $t_2 = -1.3$ eV gives $t_0 = 0.02$ eV and $\delta t = 0.02$ eV, which are used in Fig.~\ref{fig:modelband} (d).},~where both mechanisms coexist. Notably, in all cases, the system exhibits $d$-wave spin splitting characterized by a momentum-dependent sign reversal between $\Gamma-M^+$ and $\Gamma-M^-$ direction. To identify the origin of the splitting, Fig.~\ref{fig:modelband} (e) displays the orbital and atomic characters of the split bands, with the Cu $d_{x^2-y^2}$ orbital shown in the top panel and the Fe $d_{xz}+d_{yz}$ orbitals in the bottom panel. This reveals that the observed splittings predominantly arise from the Fe $d_{xz}$ and $d_{yz}$ orbitals.

{\it First-principles Realization.--} 
To realize the above models, we performed DFT calculations using VASP~\cite{Kresse1996,Hafner2008} within the generalized gradient approximation (GGA)~\cite{PBE1996}. We considered the FeSe/Bi-2201 heterostructure (space group $I4/mmm$), consisting of a single Fe layer interfaced with a single CuO$_2$ plane~\cite{wang2023prominent}, using structure A [Fig.~\ref{fig:DFT} (a)], in which the Bi-2201 slab is exfoliated at the BiO$_2$ plane. For the GGA+$U$ calculations, we employed $U_{\rm Cu}=8$~eV and $U_{\rm Fe}=0$~eV. The value of $U_{\rm Cu}$ is consistent with those commonly used for Cu $3d$ orbitals in cuprates and compensates for the tendency of GGA to underestimate the on-site Coulomb interaction, thereby providing a realistic description of the CuO$_2$ electronic structure~\cite{Anisimov1991,Liechtenstein1995,Anisimov1997}. This Hubbard $U$ should not be directly compared with the smaller interaction used in the mean-field model, where the hopping parameters already describe an effective low-energy Hamiltonian renormalized by electronic correlations. In contrast, the more itinerant Fe $3d$ states are well described with $U_{\rm Fe}=0$~eV~\cite{yamada2021multipolar}. Further computational details are provided in the Supplemental Material.

This heterostructure incorporates both AM routes. Upon optimizing the atomic positions, the two distinct Fe-Se bond lengths yield a distortion parameter $\delta_l = (l_1 - l_2)/l_2 = -0.002$ [see Fig.~\ref{fig:model} (a)]. 
The epitaxial strain from Bi-2201 (lattice constants $a = 5.388$~\AA) expands the in-plane lattice constant of FeSe from 3.76~\AA~to 3.81~\AA, inducing a magnetic state. 
The checkerboard state is a key ingredient for both AM routes. However, the magnetic ground state of monolayer FeSe is known to sensitively depend on lattice constants and the exchange-correlation functional: GGA favors a pair-checkerboard state~\cite{liu2015first, wang2015magnetic}, while r$^2$SCAN stabilizes a stripe order~\cite{myers2025stripe}. In our calculations, in addition to the above two factors, the mutual influence between layers could also modify the magnetic behavior. GGA+$U$ yields a checkerboard configuration, which has a small magnetic moment on Cu (whereas the stripe state does not). r$^2$SCAN gives a stripe state but with an excessively large Fe moment, indicating that semilocal functionals may not quantitatively capture the magnetic interactions. In view of this sensitivity, we do not attempt to resolve the ground-state controversy here. Instead, we adopt the checkerboard state as a working hypothesis, motivated by the considerations below.

Mean-field studies based on the $J_1$-$J_2$-$J_3$-$K$ spin model~\cite{glasbrenner2015effect}, where $J_1$, $J_2$, and $J_3$ are the nearest-, next-nearest-, and third-nearest-neighbor Heisenberg exchanges, respectively, and $K$ is the biquadratic exchange, predict a parameter region favorable to the checkerboard state. Scanning tunneling microscopy (STM) measurements on FeSe monolayers also suggest such a state~\cite{qiao2020fingerprint}. We therefore assume the checkerboard state and investigate the possibility of realizing AM. The structural conditions required for its stabilization remain an open question, requiring further experimental and theoretical studies beyond DFT (e.g., DFT+DMFT). Conversely, observing spin-split bands associated with AM would provide evidence for the checkerboard state in FeSe.

The electronic structure of the heterostructure shows a band splitting of 15 meV along the $M^+$-$\Gamma$-$M^-$ path [see Fig.~\ref{fig:DFT} (c)], consistent with the model predictions. To probe the impact of stronger Cu-Fe coupling in the Lieb lattice, we designed structure B [Fig.~\ref{fig:DFT} (b)] via cleaving Bi-2201 to reveal the CuO$_2$ plane. This shortens the Fe-Cu interlayer distance from 9.12 \AA\ to 4.30 \AA, thereby enhancing the coupling between layers compared to structure A. The enhanced hopping $t_0$ (from 0.5 to 5~meV) leads to a larger band splitting 25 meV, along with additional split bands emerging around the Fermi level along $M^+$-$\Gamma$-$M^-$, as shown in Fig.~\ref{fig:DFT} (d). 

For structure A, the bands around the Fermi level are predominantly contributed by Fe atoms [see Fig.~\ref{fig:DFT} (e)], and the $d_{xz}+d_{yz}$ orbitals capture most of the features relevant to the splittings [see Fig.~\ref{fig:DFT} (f)]. In addition, the CuO$_2$ layer exhibits band splitting induced by its proximity to AM FeSe (see Supplemental Material for details). Similar scenarios for the altermagnetic proximity effect have been proposed in other materials~\cite{zhu2026altermagnetic}. We also present two additional cases in the Supplemental Material: (1) an isolated FeSe monolayer, in which the structural inequivalence between the top Se1 and bottom Se2 atoms, arising from the presence of a substrate, illustrates the second mechanism; and (2) a bilayer FeSe/Bi-2212 heterostructure, which likewise forms a Lieb lattice but features different lattice constants and two Cu layers, providing a useful point of comparison.

{\it Spin conductivity.--} To explore the experimental signatures of the AM phase, we focus on the spin conductivity~\cite{gonz2021efficient}. In AM materials, anisotropic spin-polarized Fermi surfaces as those shown in Fig.~\ref{fig:modelband} generate characteristic spin currents under an electric field. The spin conductivity is described by a tensor $\sigma_{bc}^a$, where $a$ denotes the spin-polarization direction of the spin current, $b$ the flow direction of the spin current, and $c$ the direction of the applied electric field. We quantify this response by calculating the spin conductivity via the Kubo formula~\cite{gonz2021efficient,leeb2024spontaneous}:
\begin{align}
\sigma_{bc}^a = -\frac{e\pi}{N}\sum_{\mathbf{k},n,m} &A_n(\mathbf{k}, \omega)\langle u_n(\mathbf{k})|J_b(s^z)|u_m(\mathbf{k})\rangle \nonumber\\
&\times A_m(\mathbf{k}, \omega)\langle u_m(\mathbf{k})|v_c(\mathbf{k})|u_n(\mathbf{k})\rangle
\label{eq:Kubo}
\end{align}

\begin{figure}
\center
\includegraphics[width=\linewidth]{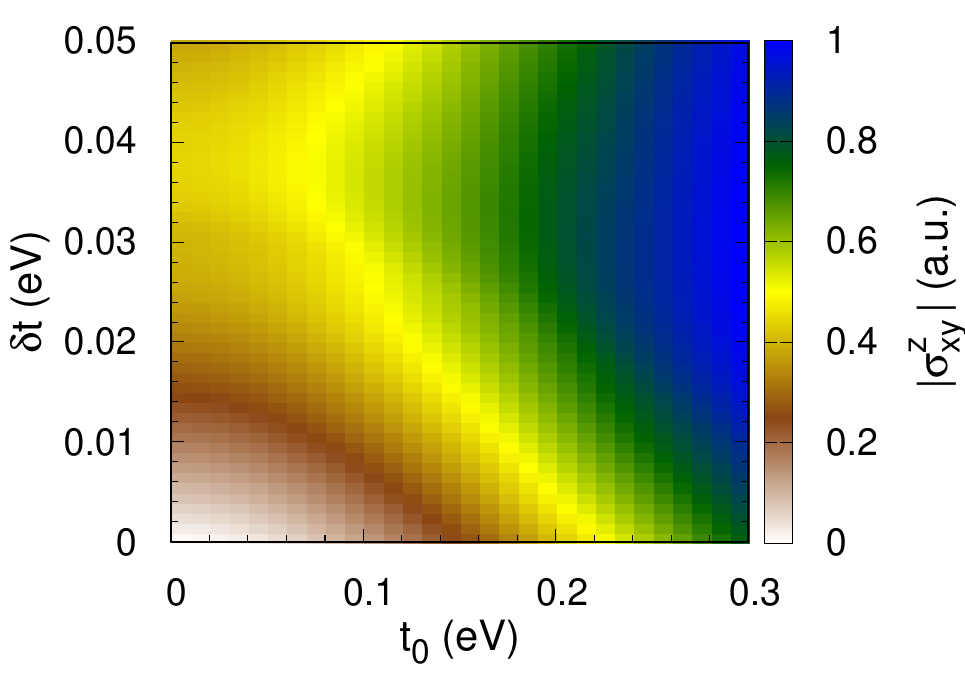}
\caption{Color plots of the transversal component of the spin conductivity $\sigma_{xy}^z$ = $\sigma_{yx}^z$, normalized to its maximum value.} 
\label{fig:spin_conduc}
\end{figure}

Here, $A_n(k, \omega) = -\frac{1}{\pi}\frac{\Gamma}{(\omega-\epsilon_n(\mathbf{k}))^2+\Gamma^2}$ is the band-resolved spectral function, where $\epsilon_n(\mathbf{k})$ is the band energy (in eV) and $\Gamma$ is the broadening parameter. $v_c = \partial h/\partial k_c$ is the velocity operator of the conduction electrons along the $c$ direction, $N$ is the total number of $\mathbf{k}$-points, and the spin current operator is defined as $J_b(s^z) = \frac{1}{2}\{s^z \otimes \mathbb{1}, v_b\}$, where $\mathbb{1}$ is the $10\times10$ identity matrix. The Hamiltonian $h(\mathbf{k})$ is the same as the one described in Eq.~(\ref{eq:10band1}).  By symmetry, the only non-zero spin conductivity components  $\sigma_{xy}$ and $\sigma_{yx}$ are equal, as shown in Fig.~\ref{fig:spin_conduc}. We compute the spin conductivity as a function of $t_0$ and $\delta t$ while keeping all other model parameters the same as in Fig.~\ref{fig:modelband} (a). Since our goal is to compare the relative effects of $t_0$ and $\delta t$ rather than to match a specific experiment, we present the spin conductivity normalized to its maximum value in arbitrary units. Other parameters are described in Ref.~\footnote{All energies ($\epsilon_n(\mathbf{k})$, $\omega$, and $\Gamma$) are in eV. We adopt atomic units with $\hbar = 1$ and set $e = 1$; thus $\sigma_{bc}^a$ is a dimensionless number in units of $e^2/\hbar$. The broadening parameter is $\Gamma = 60$ meV, chosen to ensure numerical stability across the full parameter range (bandwidth $\sim 0.5$ eV). The total number of $\mathbf{k}$-points in the summation is $N = 200 \times 200$.}. Both $t_0$ and $\delta t$ induce spin splitting and enhance the spin conductivity, with $t_0$ having the stronger effect due to its more pronounced modification of the Fermi energy. 

{\it Conclusions.-} 
In this work, we have shown that $45^\circ$-twisted FeSe/cuprate heterostructures form an effective Lieb lattice and support altermagnetism through two distinct mechanisms, with the resulting band splittings in the FeSe also transferred to the Cu layer by proximity.
Unlike previous proposals that combine an altermagnet with a conventional superconductor~\cite{heinsdorf2026proximit}, our platform consists of two potential superconducting layers, FeSe and a high-$T_c$ cuprate, with distinct pairing symmetries and critical temperatures. The proximity-induced altermagnetism may, therefore, coexist with superconductivity in both subsystems, providing a setting to study their mutual coupling. This may enable phenomena such as mixed singlet-triplet pairing~\cite{jasiewicz2026interplay}, nodal superconductivity~\cite{Fukaya2026crossed}, persistent spin currents~\cite{monkman2026persistent}, superconducting diode effects ~\cite{banerjee2024altermagnetic,sim2025pair}, and the spin-current dynamo effect~\cite{heinsdorf2026proximit, monkman2026persistent}. More broadly, the twisted FeSe/cuprate heterostructure extends the family of Fe-based altermagnets~\cite{mazin2023induced, gonzalez2026coexistence, li2026topological, cui2026surface, ding2026emergent} and will provide a unique platform for investigating the interplay between altermagnetism and high-$T_c$ superconductivity. 

{\it Acknowledgements:-}
Y. L. acknowledges support from the Alexander von Humboldt Foundation through a postdoctoral Humboldt fellowship and Shaanxi Fundamental
Science Research Project for Mathematics and Physics (Grant No. 25JSQ001). 
J.K. acknowledges support from the Deutsche Forschungsgemeinschaft (DFG, German Research Foundation) under grants TRR 360 - 492547816, KN1254/1-2, KN1254/2-1 and under Germany’s Excellence Strategy EXC-2111-390814868. P.R. and J.K. acknowledge support from the Munich Quantum Valley, which is supported by the Bavarian state government with funds from the High-tech Agenda Bayern Plus. J.K. further acknowledges support from the Imperial-TUM flagship partnership and the Keck foundation.
R.V. acknowledges support by the Deutsche Forschungsgemeinschaft (DFG, German Research Foundation) for funding through Project No. TRR 288 --- 422213477 (project A05, B05) and Project No. VA 117/23-1 --- 509751747. We acknowledge the support of the Natural Sciences and Engineering Research Council of Canada (NSERC).  This work was supported by grant 369963 from the Fonds de recherche du Québec. M. G. V. received financial support from the Canada Excellence Research Chairs Program for Topological Quantum Matter.

\bibliography{biblio}

\clearpage
\appendix
\pagestyle{empty}
\begin{figure*}
\vspace*{-1.8cm}
\hspace*{-1.6cm}%
\includegraphics[width=21cm, page = 1]{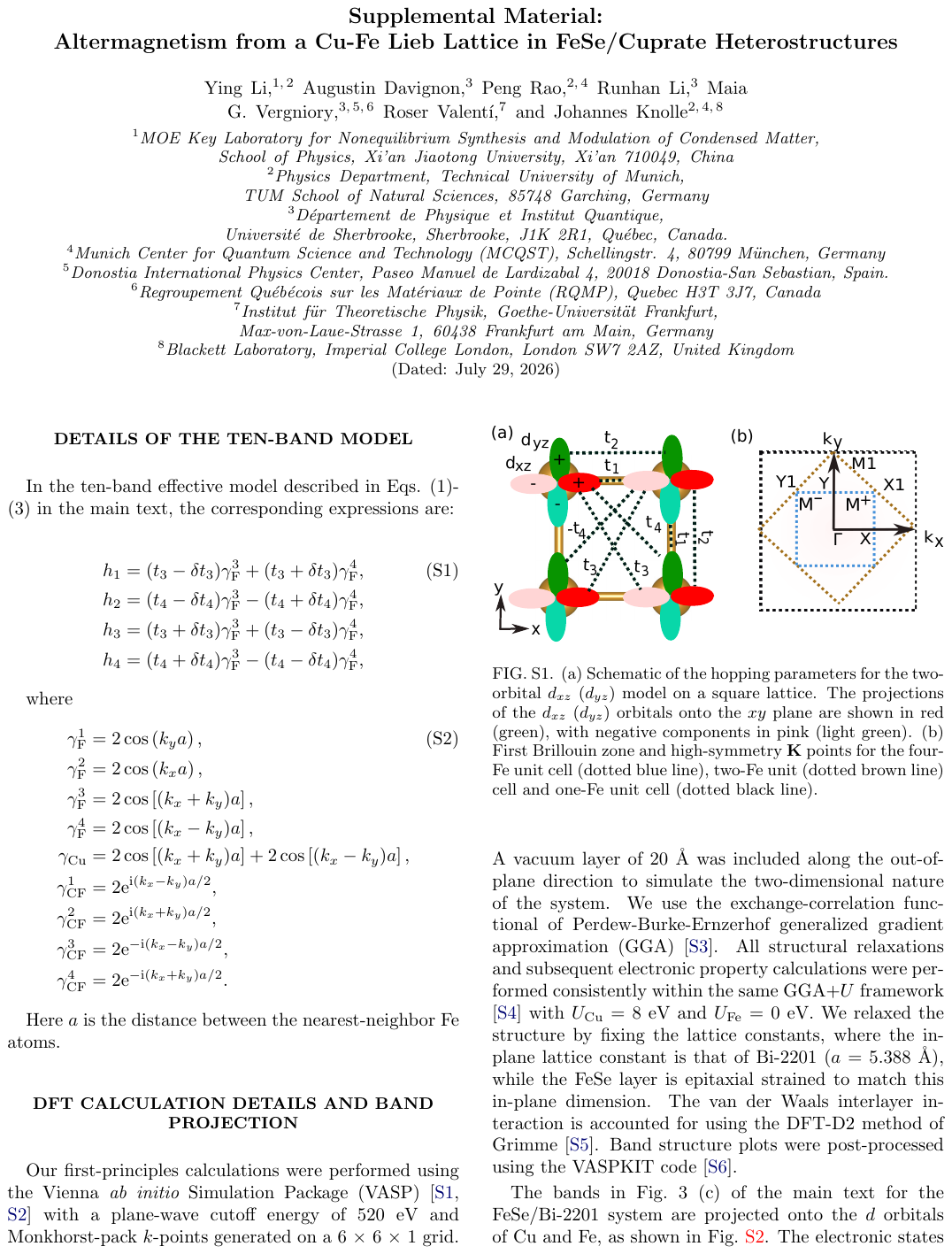}
\end{figure*}

\begin{figure*}
\vspace*{-1.8cm}
\hspace*{-1.6cm}%
\includegraphics[width=21cm, page = 2]{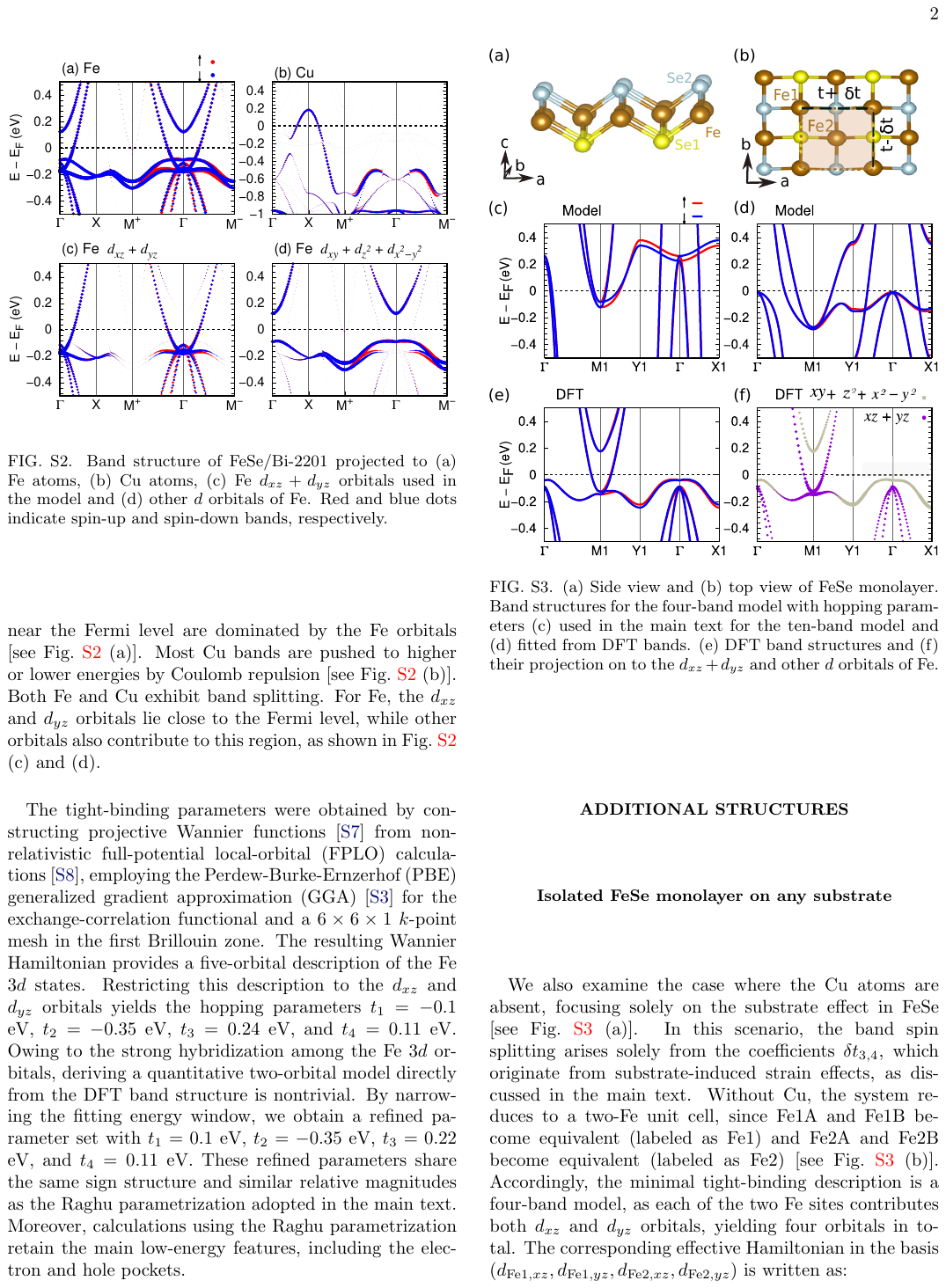}
\end{figure*}

\begin{figure*}
\vspace*{-1.8cm}
\hspace*{-1.6cm}%
\includegraphics[width=21cm, page = 3]{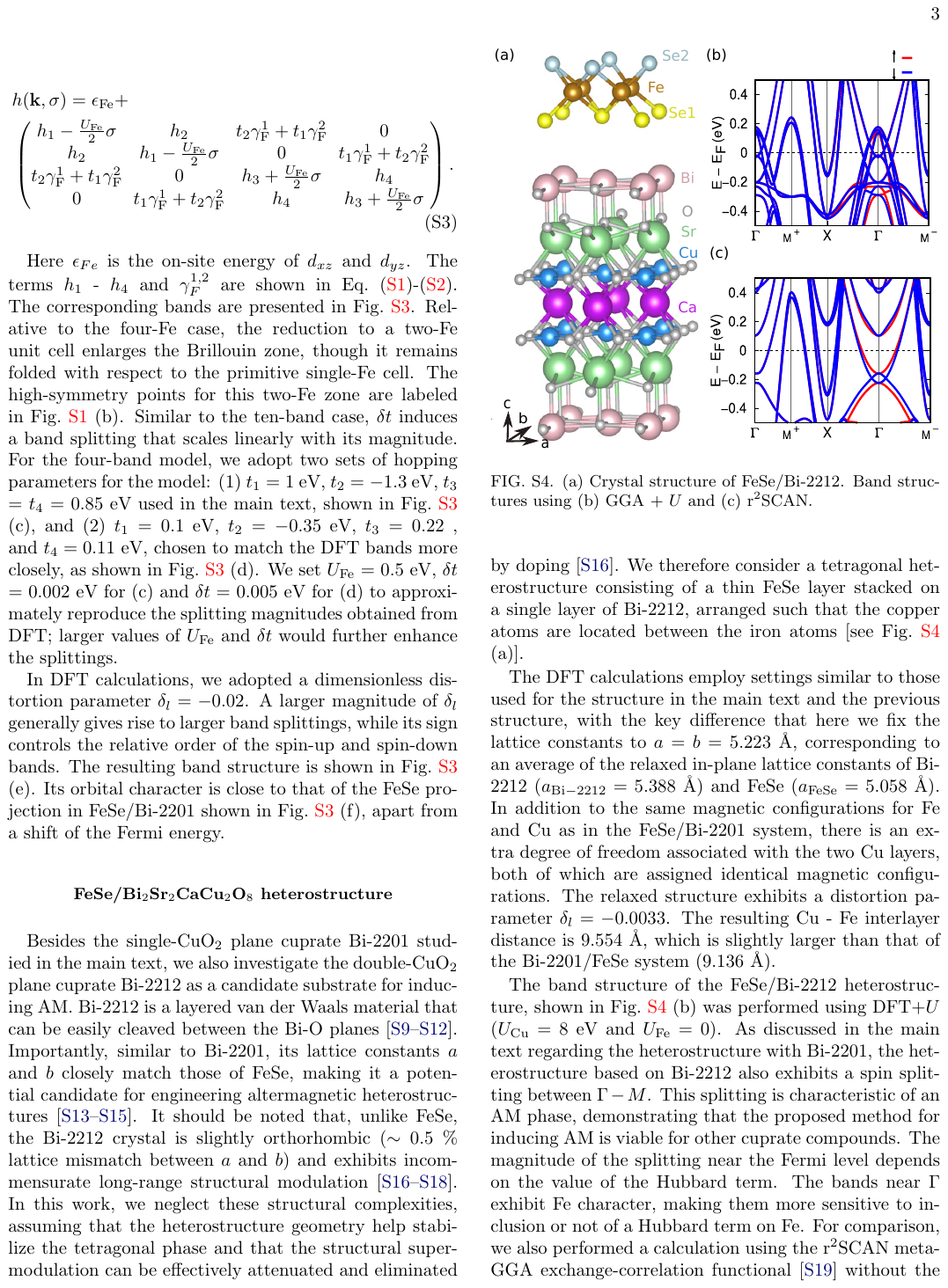}
\end{figure*}

\begin{figure*}
\vspace*{-1.8cm}
\hspace*{-1.6cm}%
\includegraphics[width=21cm, page = 4]{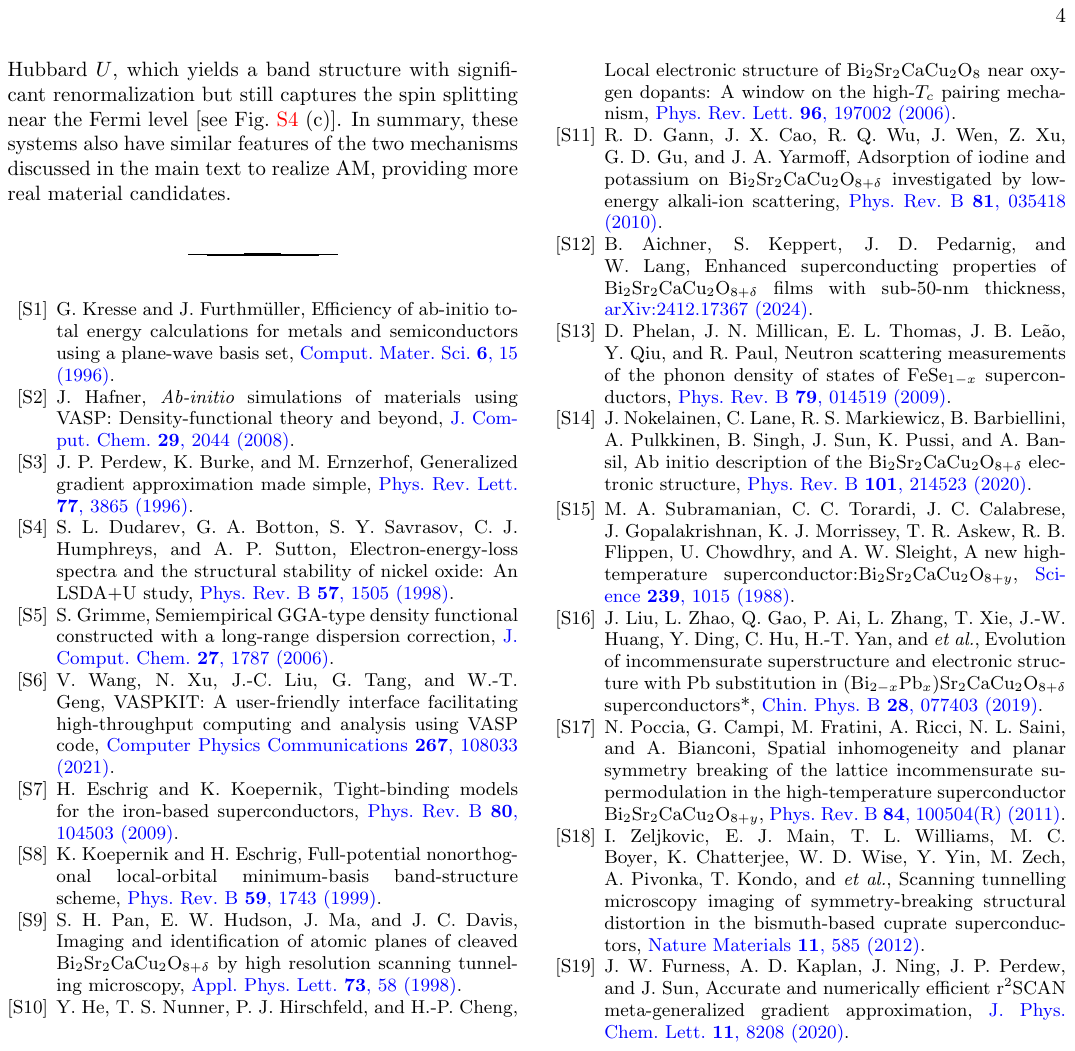}
\end{figure*}

\end{document}